\documentclass[twocolumn, trackchanges]{aastex62}

\hypersetup{linkcolor=blue,citecolor=blue,filecolor=cyan,urlcolor=magenta}

\usepackage{amsmath}
\usepackage{upgreek}

\graphicspath{{./}{figures/}}

\received{January 1, 2018}
\revised{January 7, 2018}
\accepted{\today}
\submitjournal{ApJ}

\shorttitle{KIC 12268220}
\shortauthors{Kaiming et al.}

\begin{document}

\title{KIC 12268220: A $\delta$ Scuti Pulsating Star and an Active Protohelium White Dwarf in an Eclipsing Binary System}

\correspondingauthor{Kaiming Cui}
\email{cuikaiming15@mails.ucas.ac.cn}

\author[0000-0003-1535-5587]{Kaiming Cui}
\affiliation{CAS Key Laboratory of Optical Astronomy, National Astronomical Observatories, Chinese Academy of Sciences, Beijing 100101, China}
\affiliation{School of Astronomy and Space Sciences, University of Chinese Academy of Sciences, Beijing 100049, China}

\author[0000-0002-0951-2171]{Zhao Guo}
\affiliation{Center for Exoplanets and Habitable Worlds, Department of Astronomy and Astrophysics, 525 Davey Laboratory,\\ The Pennsylvania State University, University Park, PA 16802, USA}
\affiliation{Copernicus Astronomical Center, Polish Academy of Sciences, Bartycka 18, 00-716 Warsaw, Poland}

\author{Qing Gao}
\affiliation{CAS Key Laboratory of Optical Astronomy, National Astronomical Observatories, Chinese Academy of Sciences, Beijing 100101, China}

\author{Juanjuan Ren}
\affiliation{CAS Key Laboratory of Optical Astronomy, National Astronomical Observatories, Chinese Academy of Sciences, Beijing 100101, China}

\author{Junbo Zhang}
\affiliation{CAS Key Laboratory of Optical Astronomy, National Astronomical Observatories, Chinese Academy of Sciences, Beijing 100101, China}

\author{Yutao Zhou}
\affiliation{Department of Astronomy, School of Physics, Peking University, Beijing 100871, China}
\affiliation{Kavli Institute for Astronomy and Astrophysics, Peking University, Beijing 100871, China}

\author{Jifeng Liu}
\affiliation{CAS Key Laboratory of Optical Astronomy, National Astronomical Observatories, Chinese Academy of Sciences, Beijing 100101, China}
\affiliation{School of Astronomy and Space Sciences, University of Chinese Academy of Sciences, Beijing 100049, China}
\affiliation{WHU-NAOC Joint Center for Astronomy, Wuhan University, Wuhan, Hubei 430072, China}


\begin{abstract}
We present a photometric, spectroscopic, asteroseismic, and evolutionary analysis of the Algol-type eclipsing binary KIC 12268220. 
We find the O'Connell effect and anticorrelated eclipse timing variations in the Kepler light curve, revealing the presence of large starspots. 
Radial velocities and atmospheric parameters are obtained from ground-based spectroscopic observations.
Combined with the radial velocity measurements and Gaia-derived total luminosity, our light-curve modeling yields the solution of the physical parameters for both the primary and secondary components. 
We find 14 independent frequencies arising from the $\delta$ Scuti primary, and the observed frequencies agree with the frequency range of unstable modes from nonadiabatic calculations. 
Based on the conclusion from previous literature, we run a grid of models to study the evolution process of our system.
The evolutionary tracks of our model suggest that the low-mass ($\sim 0.23\,M_\odot$) evolved secondary shows a similar evolutionary state to the R CMa-type system, which might evolve to an EL CVn system.
\end{abstract}



\section{Introduction} \label{sec:intro}

Eclipsing binary (EB) systems containing pulsating components can be used for the determination of accurate fundamental parameters.
With the photometric and spectroscopic data in the time domain, the mass and radius of the EB system can be derived accurately \citep[e.g.,][]{Southworth2005, Clausen2008}. Asteroseismic modeling can constrain the parameters of pulsators such as  $\delta$ Scuti and $\gamma$ Dor stars \citep[e.g.,][]{Chen2016, Chen2019}, which are dwarfs or subgiants located at the lower end of the classical instability strip \citep{Breger2000}. 
Many $\delta$ Scuti pulsating stars in EB systems have been discovered, especially after the Kepler mission \citep[e.g.,][]{Guo2016, Kahramans2017,Liakos2017,Gaulme2019}. 
Within the group of pulsating binaries, those of Algol-type (oEA) systems \citep{Mkrtichian2004} might experience mass transfer during their evolution.

Binary star evolution with mass transfer can generate mass-transferring or post-mass-transfer $\delta$ Scuti pulsators. One such type is the EL CVn binaries, which contain an A- or F-type primary and a low-mass helium white dwarf (WD) secondary \citep{Maxted2014, Guo2017, Zhang2017}.
Currently, more than 60 EL CVn binaries are known, with 16 being discovered through the Kepler mission \citep{Lee2018, Wang2019, Zhang2019}; however only a few have pulsation signals.

KIC 12268220 ($ K_p = 11.425 $\, mag, $\alpha$ = 19:45:57.761 $\delta$ = +50:54:21.098) was discovered as an Algol-type EB system with an orbital period of $P_\mathrm{orb} = 4.42158021\pm3.948\times10^{-6}$\,days \citep{Prvsa2011, Slawson2011}. 
According to known parameters (see Table \ref{tab: archive_parameters}), the primary star is a late-A or early-F subgiant ($T_\mathrm{eff} \sim 7800\,\mathrm{K}$, $\log g \sim 3.6$). 

In this paper, we study the activity characteristics of KIC 12268220 from its Kepler light curve and eclipse timing variations (ETV) in Section \ref{subsec: kepler}. Then, we derive the atmospheric and orbital parameters with the spectra fitting and light-curve modeling (Section \ref{subsec: spectra} and Section \ref{sec: lc modeling}). In Section \ref{sec: pulsation}, we study its pulsation properties with the nonadiabatic calculations. In Section \ref{sec: evolution}, comparing to the theoretical evolution models, we suggest KIC 12268220 would evolve to an EL CVn system. Finally, we summarize our results in Section \ref{sec: summary}.

\begin{deluxetable*}{lccc}
\tablecaption{Table of stellar parameters from archived catalogs. }
\label{tab: archive_parameters}
\tablehead{
\multicolumn{1}{l}{Parameters} & \colhead{KIC \citep[Kepler Input Catalog;][]{Brown2011}} & \colhead{Stellar17 \citep{Mathur2017}} & \colhead{Gaia DR2 \citep{Collaboration2018}}
}
\startdata
ID & 12268220 & 12268220 & 2135386791013636352 \\
$T_{\mathrm{eff}}$ (K) & 7826 &  $8026{+251 \atop -306}$  & 7835.00 \\
$\log$ g (dex) & 3.581 & $3.650{+0.510 \atop -0.090}$ & \\
$\left[ \mathrm{Fe}/ \mathrm{H} \right]$ (dex) & -0.335 & $-0.360{+0.200 \atop -0.300}$ & \\
Mass ($M_\odot$) & $1.698^*$ & $1.950{+0.238 \atop -0.510}$ & \\
Radius ($R_\odot$) & 3.494 & $3.461{+0.607 \atop -1.821}$ & 3.51  \\
Parallax (mas) &  &  & $0.7175\pm 0.0231$ \\
Distance (pc) &  & 1175.405 & $1337.7926{+42.6601\atop-40.1025}^*$
\enddata
\tablenotetext{*}{The distance of Gaia DR2 is calculated from the parallax, and the mass of KIC is derived from the $\log g$ and radius.}
\end{deluxetable*}

\section{Kepler Photometry and Spectral Analysis}\label{sec: photometry and spectra}

\subsection{Kepler Photometry}\label{subsec: kepler}
\begin{figure*}
\centering
    \includegraphics[scale=0.8]{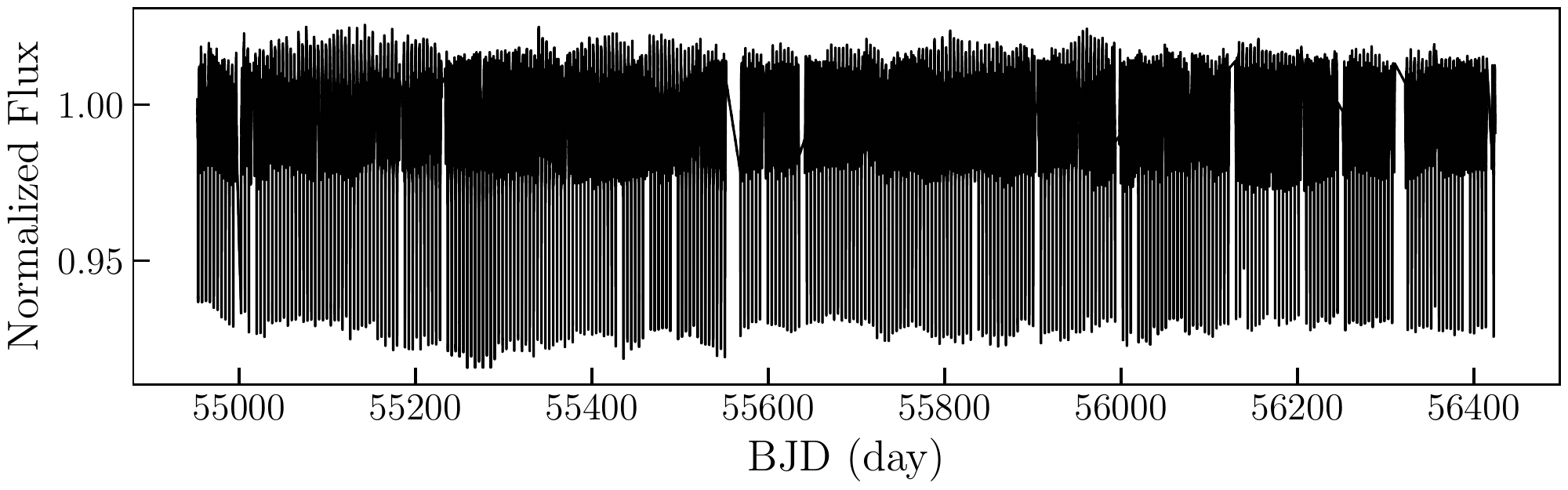}
    \includegraphics[scale=0.8]{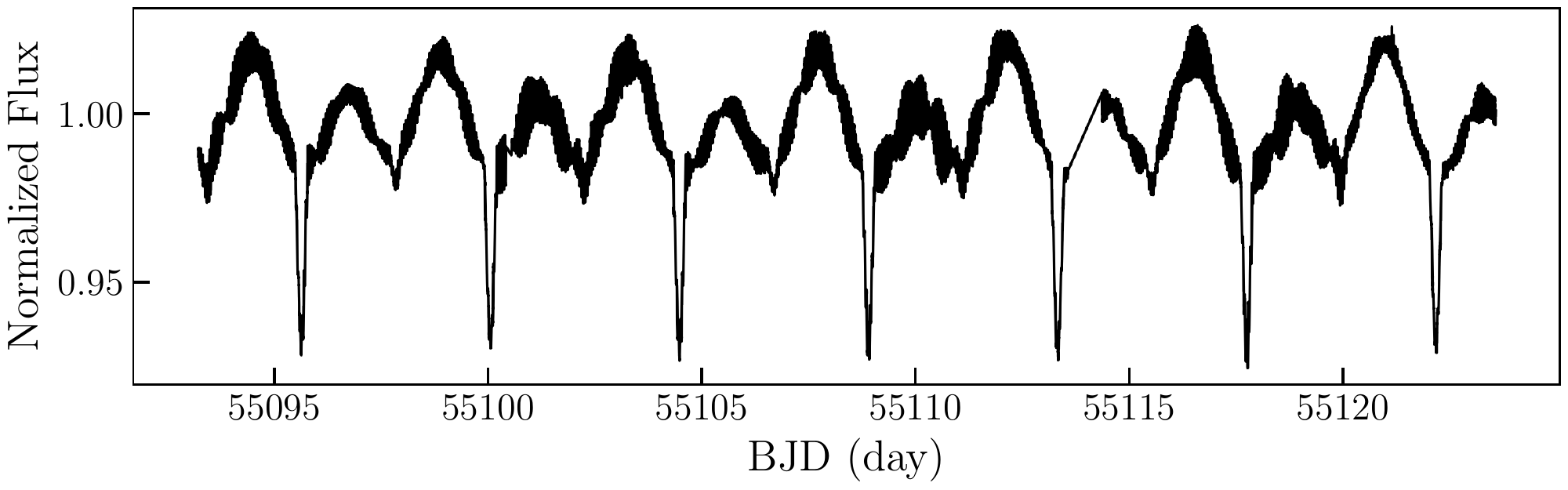}
    \caption{Detrended Kepler light curves of KIC 12268220. The top panel is the long-cadence light curve and the bottom panel is the one-month short-cadence light curve.} \label{fig: whole_lc}
\end{figure*}

KIC 12268220 was observed by Kepler from quarters 0 to 17 in the long-cadence mode (29.4 minute sampling) and one-month short-cadence (59\,s sampling) mode. The detrended and normalized light curves from the Kepler Eclipsing Binary Catalog \citep[KEBC\footnote{http://keplerebs.villanova.edu};][]{Prvsa2011, Slawson2011} are used in this work. Both long-cadence and short-cadence light curves are shown in Figure \ref{fig: whole_lc}.

From the light curve we find a strong O'Connell effect \citep[][significant flux differences at quadrature phases]{OConnell1951}, which can be clearly found in the short-cadence data (bottom panel of Figure \ref{fig: whole_lc}). 
This effect suggests the possible presence of starspots \citep[e.g.,][]{Linnell1986, Kang2004, Qian2005}. 
However, the long-lived O'Connell effect needs long lifetimes of starspots.
The lifetimes of starspots vary significantly for different spectral types of stars \citep{Giles2017}.
For solar-type stars, the lifetimes of starspots range from 10 days to a year, depending on the area of starspots \citep{Namekata2019}.
The lifetimes of cooler stars and active close binary stars (RS CVn-type stars) can be up to several years \citep[e.g.,][]{Strassmeier1994, Henry1995, Strassmeier1999a, Strassmeier1999, Giles2017}. 
Moreover, \citet{Hussain2002} found the spots on tidally locked binary systems live longer than spots on single main-sequence stars.
Since the effective temperature ratio of KIC 12268220 can be estimated with the ratio of the eclipse depth, we can infer the secondary star should be a K-type star \citep{Prvsa2011}. 
Therefore, the O'Connell effect is more likely caused by the starspots on the secondary star.

\begin{figure*}
\centering 
    \includegraphics[scale=0.7]{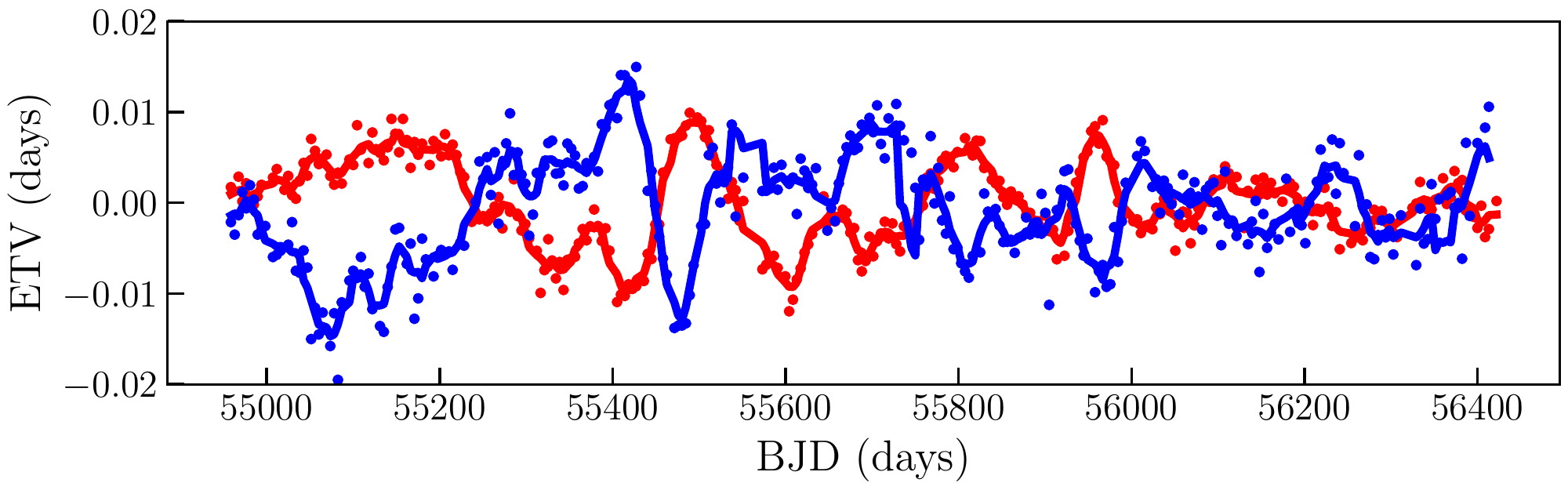}
    \caption{Anticorrelated and quasi-periodic ETV curve of the primary eclipses (red dots) and the secondary eclipses (blue dots). The red and blue lines are smoothed ETVs with a 5 point boxcar kernel.}
    \label{fig: o-c}
\end{figure*}

Additionally, the ETV results of KEBC, KIC 12268220 display an apparent variation for both the primary and secondary eclipses.
However, many secondary eclipse time variations have the same maximum or minimum values. These results may be caused by the poor fit for the secondary eclipses in their pipeline \citep{Conroy2014}. 
We fit the secondary eclipses with a second-order polynomial function to look for the time of the deepest points, and then calculate new ETVs for those secondary eclipses. Combined with the ETV results of the primary eclipses from KEBC, we find an obvious anticorrelated relation (see Figure \ref{fig: o-c}). This anticorrelation can be successfully explained by the moving of starspots, and the quasi-periodic variation of the amplitude in the ETV curve may imply the long-term evolution of the starspots \citep{Tran2013, Balaji2015}. 
However, from the simple model of \citet{Tran2013} and \citet{Balaji2015}, the sum of the colatitude and inclination angle should be less than $90^\circ$, which means there could be some polar spots in the KIC 12268220, because its inclination angle is relatively high based on its light-curve shape.
Therefore, from the evidence of long-lived starspots, polar spots, and short orbital period,
we could infer the potentially strong magnetic field of KIC 12268220 \citep{Schuessler1992, Berdyugina2005}.

\subsection{Spectral Analysis}\label{subsec: spectra}
We secured seven nights from 2018 to 2019 on the 2.16 m telescope at Xinglong Station, which is administered by National Astronomical Observatories, Chinese Academy of Sciences. 
We obtained 12 echelle spectra with the Beijing Faint Object Spectrograph and Camera (BFOSC) E9+G10 ($ R \sim 2500 $), G11 and G12.
They are three combinations of echelle and grisms with some different wavelength ranges and resolutions \citep[details in][]{Fan2016}.
After removing three spectra with signal-to-noise ratios (S/Ns) less than 30, all the spectroscopic data are reduced using the Image Reduction and Analysis Facility (IRAF) package \citep{Tody1986, Tody1993}, following the standard procedures introduced by the 11th Xinglong Observational Astrophysics Training Workshop \footnote{http://xinglong-workshop11.csp.escience.cn/dct/page/1}. 

To measure the radial velocities (RVs), we apply the template fitting method with iSpec \citep{Blanco-Cuaresma2014, Blanco-Cuaresma2019}. Before the measurements, we convert the spectrum from the air wavelength to the vacuum wavelength based on the method of \citet{Birch1994}.
According to the parameters in Table \ref{tab: archive_parameters}, we choose $T_\mathrm{eff} = 7800$ \,K, $\log g = 3.6$, and $\left[ \mathrm{Fe}/ \mathrm{H}\right] = -0.3$ as the initial values to generate a template spectrum. Then, we apply the cross correlation algorithm to find the best fitted RV for each observed spectrum. iSpec can be used to analyze the double-lined spectroscopic binary. However, limited by the resolution and the luminosity difference between the primary and secondary stars, we can only measure the RVs from the primary star. Also, although most of the spectra were observed consecutively in 2018, two spectra were obtained in 2019. Thus, we add the barycentric velocity correction for each RV. 
The RV results are listed in Table \ref{tab: rv_results}. Our measurement uncertainties correlate well with the wavelength calibration results of the BFOSC \citep[in preparation]{junbo}.

\begin{deluxetable}{ccccc}
\tablecaption{Table of RV measurements of the primary star.\label{tab: rv_results}}
\tablehead{\colhead{Date}&\colhead{Orbital}&\colhead{RV}&\colhead{RV Error}&\colhead{Instrument}\\
\colhead{(MJD)}&\colhead{Phase}&\colhead{(km s$^{-1}$)}&\colhead{(km s$^{-1}$)}&}
\startdata
58763.508            & 0.1665                & 71.47                & 5.00     & E9+G12           \\
58417.576            & 0.9293                & 47.87                & 4.33     & E9+G10           \\
58419.609            & 0.3891                & 70.48                & 3.98     & E9+G10           \\
58419.630            & 0.3939                & 69.09                & 4.39     & E9+G10           \\
58420.648            & 0.6242                & 50.36                & 4.84     & E9+G10           \\
58420.620            & 0.6179                & 52.39                & 4.12     & E9+G10           \\
58762.644            & 0.9712                & 56.83                & 7.66     & E9+G11           \\
58418.568 & 0.1537 & 68.80 & 8.63 & E9+G10 \\
58385.527 & 0.6810 & 43.32 & 5.21 & E9+G10 \\
\enddata
\end{deluxetable}

In order to obtain the atmospheric parameters, we combine three best-quality spectra of E9+G10 to increase the S/N.
The combined spectrum is also resampled to keep the wavelength steps consistent with the raw spectrum.
Then, the atmospheric parameters are derived by using iSpec with the synthetic spectral fitting technique. iSpec implements some commonly used synthetic models and atomic line lists. In this work, considering a higher $T_{\mathrm{eff}}$ ($ >7000$\,K) and for the wavelength of the spectra, we choose the Vienna Atomic Line Data Base (VALD) line lists \citep{Ryabchikova2015}, the SPECTRUM code \citep{Gray1994}, the Grevesse 2007 solar abundances \citep{Grevesse2007}, and the ATLAS9 Castelli atmosphere library \citep{Kurucz2005}. 

\begin{figure}
\centering 
    \includegraphics[width=\columnwidth]{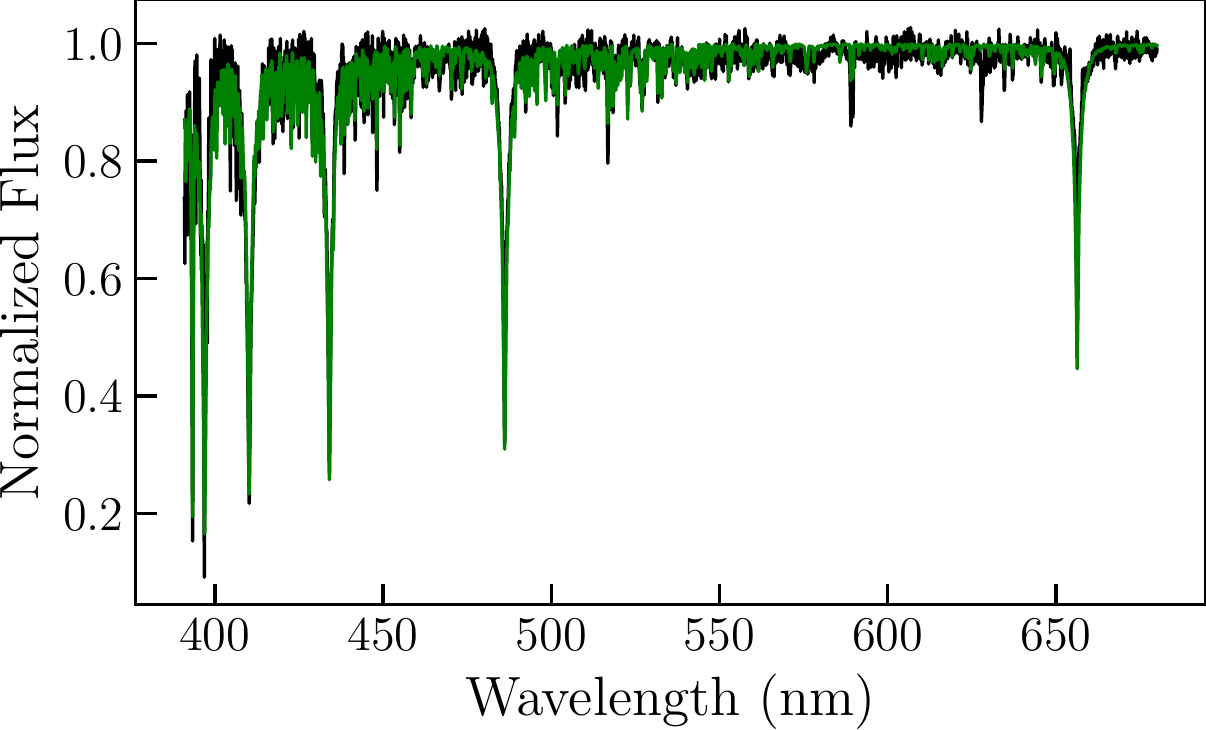}
    \caption{Observed composite spectrum (black) and the fitted model spectrum from iSpec (green).} 
    \label{fig: fitted_spectrum}
\end{figure}

The initial parameters are set to the same values as the radial velocity calculation, and the resolution is fixed at 2500. 
Because of the relatively short orbital period, synchronous rotation could be expected. Thus, we specify the $v_\mathrm{rot}\sin i$ at $37\ \mathrm{km}\,\mathrm{s}^{-1}$ with the estimated radius $R \approx 3.5\,R_\odot$ and inclination angle $i \approx 70^\circ$.
iSpec applies the Levenberg--Marquardt algorithm to fit the spectrum, and the iteration stops when the ftol (relative error desired in the sum of squares) or xtol (relative error desired in the approximate solution) less than $10^{-10}$. The corresponding errors are calculated from the covariance matrix.
However, our internal errors are probably underestimated. Thus a more robust error determination is needed. Because of the slightly higher resolution, the BFOSC was often used as the follow-up confirmation and calibration of the Large Sky Area Multi-Object Fiber Spectroscopic Telescope \footnote{http://www.lamost.org} \citep[LAMOST;][]{Fan2016}. 
Therefore, we choose the mean errors of the late A-type subgiants from the LAMOST spectrum as the estimated errors. After we consider the statistic errors, the best-fitting results and their errors are listed in Table \ref{tab: spec_atmo_para}.
The observed composite spectrum and the model spectrum are shown in Figure \ref{fig: fitted_spectrum}. The observed spectrum is dominated by the strong hydrogen Balmer series, without obvious peculiar metallic-line weakness or enhancement (e.g., \ion{Ca}{2} K, \ion{Si}{2}, \ion{Cr}{2}, and \ion{Sr}{2}), which matches the model spectrum well.

\begin{deluxetable}{lc}
\tablecaption{Table of atmospheric parameters from spectra fitting.\label{tab: spec_atmo_para}}
\tablehead{\multicolumn{1}{l}{Parameters (Units)}&\colhead{Fitted Results}}
\startdata
$T_{\mathrm{eff}}$ (K)               & $7843\pm105$ \\
$\log$ g (dex)                       & $3.75\pm0.22$     \\
$\left[\mathrm{M}/\mathrm{H}\right]$ (dex) & $-0.29\pm0.1$    \\
$ \left[ \alpha / \mathrm{Fe} \right] $ (dex) & $0.12\pm0.1$  \\
Microturbulence velocity (km\,$\mathrm{s}^{-1}$) &  $2.48$  \\
Macroturbulence velocity (km\,$\mathrm{s}^{-1}$) &  $28.57$ \\
$ v \sin i$ (km\,$\mathrm{s}^{-1}$)  &     $37$   \\
Resolution                           &     2500     \\
\enddata
\tablecomments{The microturbulence velocity and macroturbulence velocity are adopted by an empirical relation considering the effective temperature, surface gravity, and metallicity. The relation was constructed by the Gaia-ESO Survey.}
\end{deluxetable}


\section{Light-curve Modeling}\label{sec: lc modeling}
Since we have the atmospheric parameters and RV measurements of the primary star, we can combine with the light curve to constrain the parameters of the invisible secondary star. To do so, we use the PHOEBE \citep{Prsa2005}, which is based on the Wilson--Devinney \citep{Wilson1971, Wilson1979, Wilson1990, Wilson2014} code, to fit the phase folded light curve in the semidetached mode. However, because the variation of starspots dominates the dispersion of the folded light curve, we use the Savitzky--Golay filter \citep{Savitzky1964} to obtain a smoothed version of the folded light curve, and our fitting result is based on these light curves.

The effective temperature of the primary star is fixed to the value from our spectra fitting results (7843\,K). 
We also set $e=0$ with the assumption of circular orbit based on the short orbital period and the phase difference between the two eclipses of the folded light curve \citep{Zhang2018}. The albedos are set to 1.0 and 0.5 for primary and secondary stars, respectively \citep{Lucy1967, Rucinski1969}; the gravity brightening coefficients are adopted to 1.0 and 0.32, respectively.

In addition, as we discussed in Section \ref{subsec: kepler}, the O'Connell effect, and the anticorrelated ETV curves evidently show the existence of starspots in KIC 12268220, and the starspots could seriously affect the results of our fitting. The location, size, and temperature of a starspot are usually strongly correlated. To solve this problem, we apply some prior knowledge to the parameters of starspots.
Firstly, although there are some A-type stars showing the spot-like features in the light curves \citep[e.g.,][]{Balona2013, Balona2017}, because of the deeper convective envelopes of the late-type stars, starspots are more likely to appear on these types of stars \citep{McQuillan2014}. We also exclude the possibility that the spots are caused by chemically abundant inhomogeneities \citep[classified as ACV variables;][]{Bernhard2015} through our spectrum observation. Therefore, the starspots should be on the secondary star. Second, the ratio of starspot temperature to stellar effective temperature could be derived from \citet{Berdyugina2005} and \citet{Maehara2017} as Equation \ref{eq: tspot}
\begin{equation}\label{eq: tspot}
    T_\mathrm{spot} / T_\mathrm{star} = 1 - 3.58\times10^{-5} T_\mathrm{star} - 0.249 + 808 / T_\mathrm{star}.
\end{equation}
Thirdly, although there could be multiple starspots at different locations, we only add one minimal starspot to prevent overfitting. To achieve this, we choose the colatitude of the starspot equal to the inclination angle, which means the center of the starspot is facing the line of sight. 
During our fitting process, the temperature ratio of the starspot and the colatitude are calculated after each iteration. Finally, with applying the differential corrections routine, we obtain a local minimum of the parameters as an initial guess.

However, without a secondary RV curve, there is still a lot of degeneracy. Since many eclipsing binaries can be used to estimate distance accurately \citep[e.g.,][]{Guinan1998, Pietrzynski2013}, to remove this degeneracy, we use the parallax from Gaia DR2 as a piece of extra information, although the errors would be significantly large. Because the flux of Kepler light curves are not calibrated, the model cannot derive the distance directly. However, the absolute luminosity of the binary system can be estimated from the Gaia DR2 parallax. To do so, the absolute magnitude can be calculated using $M_V = m_V - 5(\log d - 1) - A_V$, where the $m_V = 11.471 \pm 0.02 $ is adopted from \citet{Everett2012} and the $A_V = 0.155 \pm0.062 $ is calculated based on the $E(B-V) = 0.05 \pm 0.02 $ from the 3D dust map \citep{Green2018,Green2019} and $ R_V = 3.1 $. The distance $ d = 1340.5\pm42$\,pc is obtained from \citet{Bailer-Jones2018}. The logarithmic absolute luminosity relative to the Sun can be estimated from $\log L/ L_\odot = -0.4(M_V + \mathrm{BC}_V - M_\mathrm{bol\odot})$, where the $M_\mathrm{bol\odot}$ is 4.74 \citep{Mamajek2015}, and the $\mathrm{BC}_V$ is the bolometric correction of $V$ band. Since the luminosity of the invisible secondary star is significantly smaller than the primary star, the $ \mathrm{BC}_V = 0.05\pm0.02 $ is interpolated from the MIST bolometric correction grids \citep{Paxton2011, Paxton2013, Paxton2015, Choi2016, Dotter2016} with our fitted atmospheric parameters. Finally, the total luminosity is $ \log L/ L_\odot = 1.60\pm0.04 $.

Then, we apply a Markov Chain Monte Carlo (MCMC) sampler to explore the posterior distribution of the binary parameters. Using MCMC to sample the posterior probability distribution is quite common to obtain more robust results of many binary systems \citep[e.g.][]{Schmid2015, Hambleton2018, IglesiasMarzoa2019, Mahadevan2019}. Our MCMC sampler is based on the emcee package \citep{ForemanMackey2013}, which is an affine invariant version of the MCMC method \citep{Goodman2010}. Because the computational load is heavy for the MCMC method, we choose the 2015 version of the  Wilson--Devinney LC code to generate model light curves and RV curves. Nine parameters are free in our sampling: the effective temperature of the secondary star, $T_\mathrm{eff,2}$; the semi-major axis, sma; the mass ratio, $q$; the inclination angle, $i$; the primary star surface potential, $\Omega_1$; the center-of-mass velocity, vga; the passband luminosity of the primary star, HLA; the longitude of the starspot, xlong; and the radius of the starspot, radsp. 
The colatitude of the starspot is set to the inclination angle, and according to Equation \ref{eq: tspot}, the temperature ratio of the starspot is restricted by the $T_\mathrm{eff,2}$.

\begin{figure*}
\centering 
    \includegraphics[width=\textwidth]{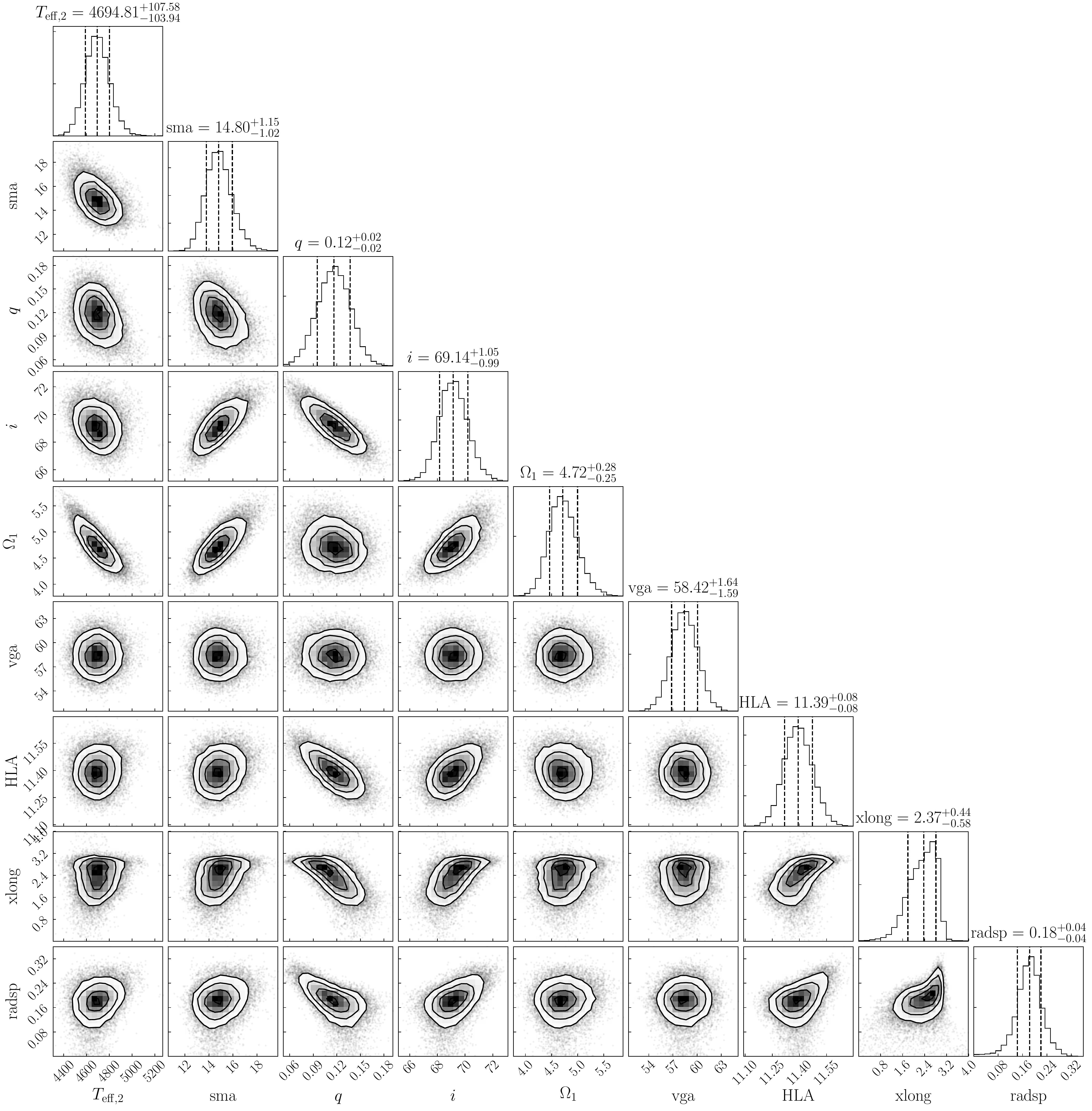}
    \caption{The marginalized posterior probability distributions of the binary star parameters. In the one-dimensional distributions along the diagonal, the 16th, 50th, and 84th percentile are indicated by the vertical dashed lines.} 
    \label{fig: corner}
\end{figure*}

Our likelihood function is written as
\begin{equation}
    \ln p(D|\Theta) = -\frac{1}{2} \left( \chi_\mathrm{LC}^2 + \chi_\mathrm{RV}^2 + \chi_\mathrm{Lum}^2 \right),
\end{equation}
where the $\chi_\mathrm{Lum}^2$ is calculated from the Gaia-derived total luminosity and the sum of model luminosity of the primary and secondary stars. The prior distribution is a uniform distribution for each parameter, and the ranges of the prior distributions are large enough to cover reasonable models. The initial parameters are obtained from the PHOEBE results. The number of walkers is 128, and to ensure convergence, we choose 30 times of the integrated autocorrelation time as the ``burn-in'' steps. After ``burnt-in,'' more than 50,000 steps are restarted, and we also thin the chains with the autocorrelation time to reduce autocorrelation. Then, we can derive the final parameters and uncertainties from their marginalized posterior probability distributions. As shown in Figure \ref{fig: corner}, for each parameter, we adopt the median value as the best-fitting value, and the 16th and 84th percentiles as the upper and lower uncertainties. Moreover, the final values and errors of masses, radius, loggs, passband luminosity ratio, colatitude of starspot, and temperature ratio of the starspot are also derived from their corresponding distributions (calculated after each iteration).

\begin{figure*}
    \centering
    \includegraphics[scale=0.65]{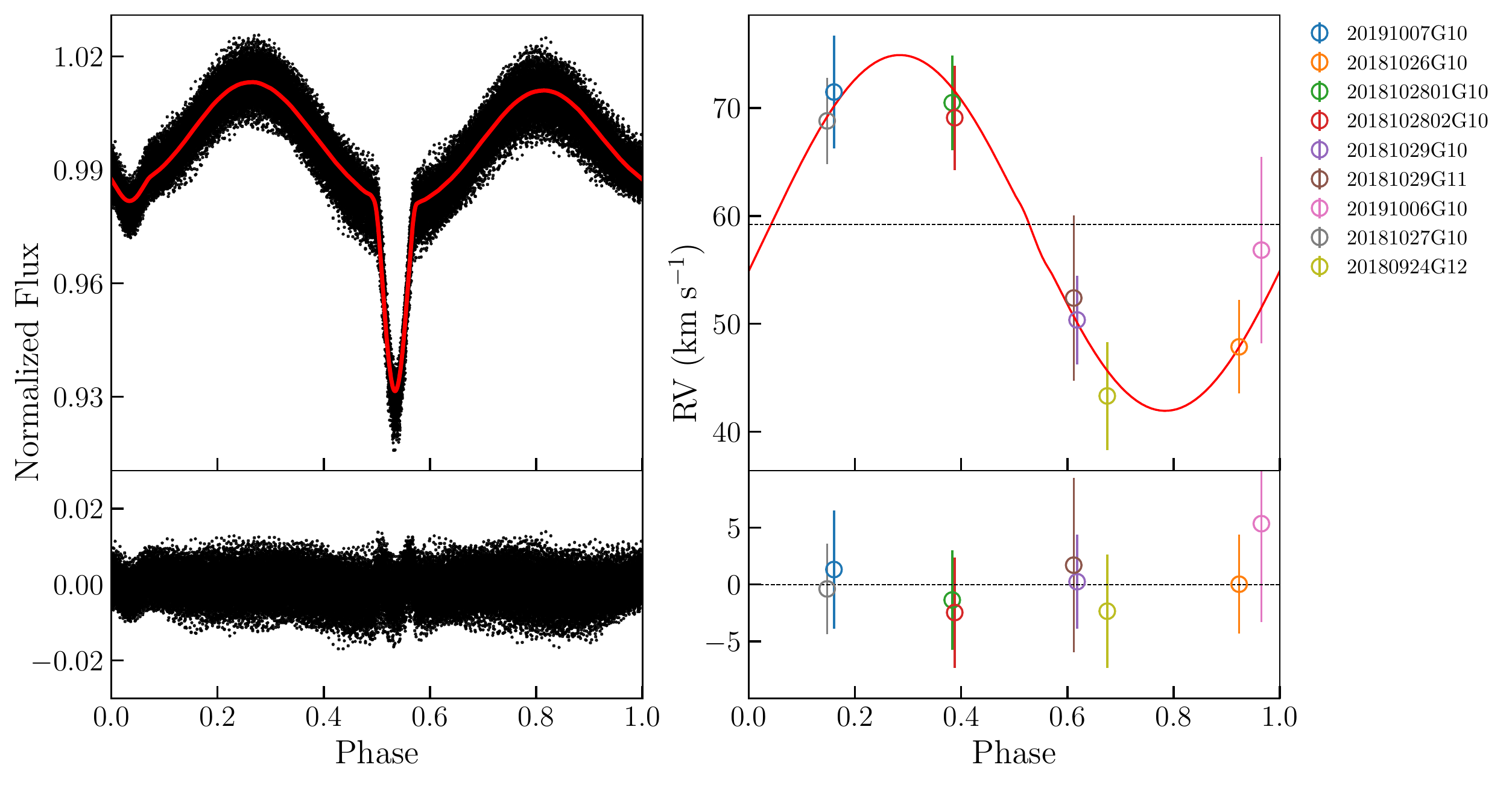}
    \caption{Left panel: phase folded light curve (black dots) and the fitted light curve (red line) with residuals in the lower panel. Right panel: the measured RVs with errors (points in color) and the best-fitting RV curve (red line).}
    \label{fig: wd_fitted}
\end{figure*}

The light-curve modeling results are given in Table \ref{tab: wd_results}, and the values without errors are the fixed parameters. 
The observed Kepler light curve and RV curve with the best-fitting models are shown
in Figure \ref{fig: wd_fitted}. The randomly distributed residual shows our model is a good solution of the light curve and RV curve.

\begin{deluxetable}{lcc}
\tablecaption{Table of MCMC results.\label{tab: wd_results}}
\tablehead{\multicolumn{1}{l}{Parameters (Units)}&\colhead{Primary}&\colhead{Secondary}}
\startdata
$T_0$ (JD)                                    & \multicolumn{2}{c}{2400000.5}                          \\
Period (days)                                  & \multicolumn{2}{c}{4.421580}                                \\
Mass ratio $q$ ($M_2 / M_1$)                     & \multicolumn{2}{c}{$0.12\pm0.02$}                                \\
Orbital eccentricity $e$                         & \multicolumn{2}{c}{0.0}                                     \\
Orbital inclination $i$ (deg)                          & \multicolumn{2}{c}{$69.14^{+1.05}_{-0.99}$}                               \\
Semi-major axis sma ($R_\odot$)                   & \multicolumn{2}{c}{$14.80^{+1.15}_{-1.02}$}                               \\
Center of mass velocity (km s$^{-1}$) & \multicolumn{2}{c}{$58.42^{+1.64}_{-1.59}$}                               \\
Primary star surface potential $\Omega_1$ & \multicolumn{2}{c}{$4.72^{+0.28}_{-0.25}$}                               \\
$L_1/ ( L_1 + L_2)_{Kp}$ & \multicolumn{2}{c}{$0.91\pm0.01$}                               \\
Gravity brightening                            & 1.0                           & 0.32                          \\
Bolometric albedo                              & 1.0                           & 0.5                           \\
$T_{\mathrm{eff}} (K)$                             & 7843                        & $4695^{+108}_{-104}$                          \\
$M (M_\odot)$                                  & $1.99^{+0.52}_{-0.40}$                    & $0.23\pm0.05$                      \\
$R (R_\odot)$                                  & $3.23\pm0.15$                    & $3.17\pm0.23$                      \\
$\log g$                                       & $3.72\pm0.08$                   & $2.80\pm0.03$                      \\ 
Luminosity ($\log L/L_\odot$)                                       & $1.54\pm0.04$                    & $0.64\pm0.04$                      \\ \hline
\multicolumn{3}{l}{Spot parameters:}                                                                         \\
Colatitude (rad)                               & \multicolumn{2}{c}{$1.21\pm0.02$}                                  \\
Longitude (rad)                                & \multicolumn{2}{c}{$2.37^{+0.44}_{-0.58}$}                                  \\
Radius (rad)                                   & \multicolumn{2}{c}{$0.18\pm0.04$}                                  \\
$T_\mathrm{spot}/T_\mathrm{local}$                               & \multicolumn{2}{c}{$0.76\pm0.01$}                                  \\
\enddata
\end{deluxetable}

\section{Pulsation Analysis}\label{sec: pulsation}
\begin{figure*}
\centering
    \includegraphics[scale=0.65]{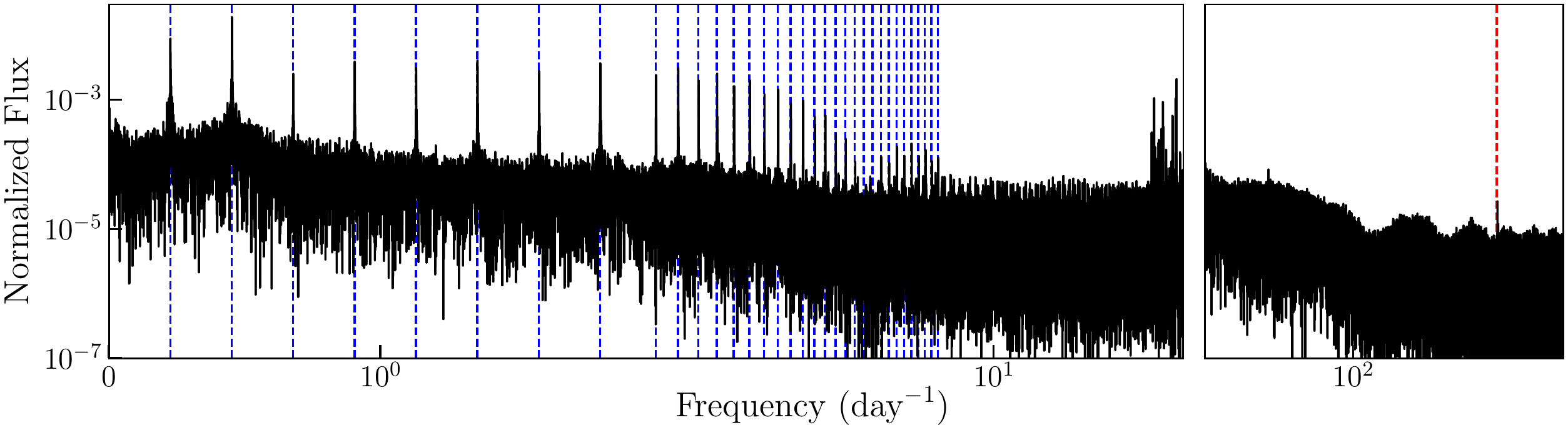}
    \includegraphics[scale=0.65]{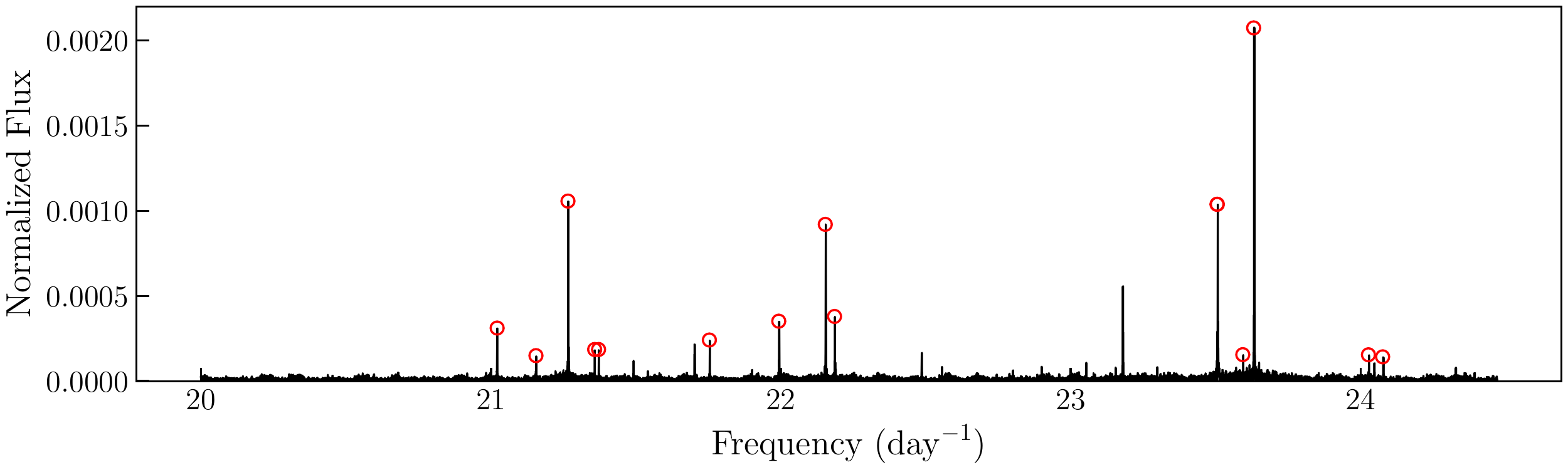}
    \caption{Fourier amplitude spectrum of the light curves. The top panel shows the overview of the amplitude spectrum in a log scale. The top left panel is calculated from the long-cadence light curve, and the dashed blue lines indicate the harmonics of the orbital frequency. The top right panel is made by the short-cadence light curve from the Nyquist frequency of long-cadence to the Nyquist frequency of short-cadence, the dashed red line is the 489.36 $\mathrm{day}^{-1}$, which is one of the spurious frequencies in short-cadence data \citep{VanCleve2016}. The bottom panel is the zoomed-in of the pulsational frequency range in linear scale, and the open red circles indicate the independent frequencies.} 
    \label{fig: freq_spectrum}
\end{figure*}

From the short-cadence light curve, we can clearly see the effect of pulsation signals on the shoulders between the two eclipses. As shown in the Fourier amplitude spectrum (see Figure \ref{fig: freq_spectrum}), the low-frequency region ($ f \lesssim 5\, \mathrm{day}^{-1}$) is dominated by the orbital frequency ($f_\mathrm{orb} = 0.22616\, \mathrm{day}^{-1}$) and its harmonics. 
The high frequency region shows typical $\delta$ Scuti pulsations with frequencies ranging from 20 to $\sim 24\,\mathrm{day}^{-1}$. Indeed, our spectroscopic parameters of the primary pin point the star in the $\delta$ Scuti instability strip (see Fig. \ref{fig: evolution}).

To investigate the pulsation properties, we apply the SigSpec \citep{Reegen2007} to the 4\,yr long-cadence light curve after removing the modeled EB light curve. Because the short-cadence data shows no significant frequencies higher than the Nyquist frequency (24.510 $\mathrm{day}^{-1}$), we calculate the significant frequencies from 0 to the Nyquist frequency. SigSpec performs a prewhitening procedure for a given light curve. The prewhitening method calculates the Discrete Fourier Transform and fits the signal with a sinusoidal of variable amplitude and phase, then iteratively subtracts the fitted light curve from the previous light curve. 
In each iteration, SigSpec calculates a $sig$ (spectral significance), which is defined as the logarithm of the inverse false alarm probability, and the false alarm probability shows that the probability of a peak is caused by pure noise in a non-equidistantly spaced data set. With the Equation (31) of \citet{Reegen2007}, $sig$ could be conveniently converted to the $ \mathrm{S/N} $. In our case, the procedure stops when $ sig < 5.46$, which is approximately equivalent to the empirical criterion: $ \mathrm{S/N} < 4 $.
After the prewhitening, besides the low-frequencies caused by the starspots and imperfect EB fitting, 19 $\delta$ Scuti frequencies are extracted and listed in Table \ref{tab: freqs}. Following the method introduced by \citet{Kallinger2008}, the errors of the frequencies, amplitudes, and phases are calculated based on their $sig$.

Because of the nonlinear effect, $\delta$ Scuti stars often show combination frequencies, we search for those combinations by computing
\begin{equation}
    \lvert f_i - \left( n f_j + m f_k \right) \rvert < \varepsilon,
\end{equation}
where $n$ and $m$ are integers (1,2,3), and $\varepsilon$ is the Rayleigh resolution ($\varepsilon = 1/\Delta T \approx 0.00068\, \mathrm{day}^{-1} $; $\Delta T \approx 1470.46\, \mathrm{days} $). If the difference between two frequencies is less than the Rayleigh resolution, the two frequencies are indistinguishable \citep{Papics2012}; and if a frequency could be combined by two parent frequencies with larger amplitudes, it is marked in Table \ref{tab: freqs}. Finally, we obtain 14 independent frequencies.

\begin{deluxetable*}{cCCCCC}
\tablecaption{$\delta$ Scuti oscillation frequencies.\label{tab: freqs}}
\tablehead{\colhead{ID}& \colhead{Frequency ($\mathrm{day}^{-1}$)} & \colhead{Amplitude (Normalized Flux)} & \colhead{Phase (rad/$2\uppi$)} & \colhead{$sig$} & \colhead{Notes}}
\startdata
$f_0$  & 23.631048(16)    & 0.001983(46)             & 0.759(11)    & 1844.86 &  \cdots                   \\
$f_1$  & 21.265576(29)    & 0.001056(45)             & 0.257(20)    & 549.76  &  \cdots                   \\
$f_2$  & 23.505764(29)   & 0.001593(67)              & 0.375(20)    & 560.84  &  \cdots                   \\
$f_3$  & 22.153237(32) & 0.000686(32)                & 0.769(22)    & 455.82  &  \cdots                   \\
$f_4$  & 23.178657(50)    & 0.000576(43)           & 0.440(34)      & 183.56  & f_0 - 2f_\mathrm{orb}   \\
$f_5$  & 22.185221(78)   & 0.000361(41)        & 0.267(53)    & 75.92   &  \cdots                   \\
$f_6$  & 21.992776(79)              & 0.000392(46)                    & 0.438(54)    & 73.44   &  \cdots                   \\
$f_7$  & 21.021237(96)              & 0.000294(42)                   & 0.567(66)    & 49.80   &  \cdots                   \\
$f_8$  & 21.753399(108)              & 0.000256(41)                   & 0.193(74)    & 39.59   &  \cdots                   \\
$f_9$  & 21.701051(130)             & 0.000208(40)                    & 0.354(89)    & 27.23   &  f_3 - 2f_\mathrm{orb}  \\
$f_{10}$ & 21.371163(163)              & 0.000174(42)                   & 0.882(112)    & 17.46   & \cdots                     \\
$f_{11}$ & 23.593506(163)              & 0.000170(41)                    & 0.750(112)    & 17.40   &  \cdots                    \\
$f_{12}$ & 21.357659(169)              & 0.000165(41)                    & 0.025(116)    & 16.11   &  \cdots                   \\
$f_{13}$ & 21.155123(179)             & 0.000157(41)                    & 0.196(123)    & 14.43   &  \cdots                   \\
$f_{14}$ & 22.485215(179)              & 0.000157(41)                    & 0.001(123)    & 14.35   &  2f_8 - f_7             \\
$f_{15}$ & 24.075994(193)             & 0.000146(41)                    & 0.609(132)    & 12.47   &  \cdots                   \\
$f_{16}$ & 23.504488(220)              & 0.000137(44)                   & 0.042(150)    & 9.59    &  \cdots                   \\
$f_{17}$ & 21.490928(232)              & 0.000120(41)                    & 0.211(159)    & 8.62    &  3f_3 - 2f_{14}           \\
$f_{18}$ & 23.051902(236)             & 0.000118(41)                    & 0.153(162)    & 8.29    & f_{16} - 2f_\mathrm{orb} \\ 
$f_{19}$ & 24.026986(245)             & 0.000114(41)                    & 0.020(168)    & 7.69    & \cdots 
\enddata
\end{deluxetable*}

\begin{figure*}
    \centering
    \includegraphics[scale=0.75]{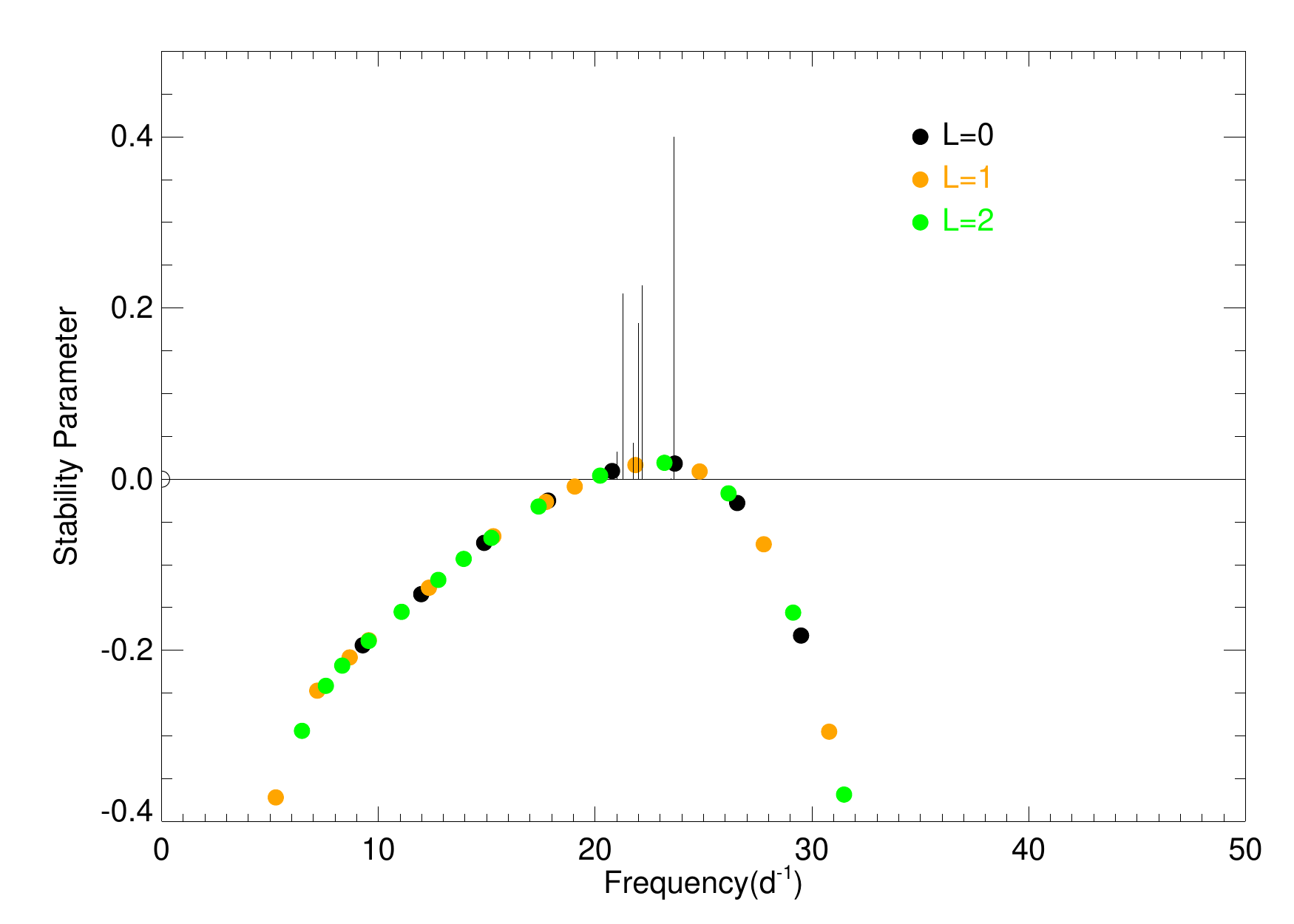}
    \caption{Stability parameter for p-modes of the model. These unstable p-modes have comparable frequencies with the observed oscillations in the Fourier spectrum (scaled and overplotted.)}
    \label{fig: etaplot}
\end{figure*}

In addition, although the pulsation properties of the primary may differ from the results of single star evolution due to the mass transfer in a binary system, we also check the frequency range of unstable modes with the non-adiabic calculation. To do so, we calculated the non-adiabaic eigen-functions and eigen-frequencies with the Dziembowski's oscillation code \citep{Dziembowski1971, Dziembowski1977} for a stellar model representing the observed parameters.
The stellar model from the MESA evolution code  \citep[][]{Paxton2011,Paxton2013,Paxton2015,Paxton2018} has the following parameters: $M=2.17\, M_\odot$, $R=3.11\,R_\odot$, $Z=0.01$.
The mass and radius are essentially within 1$\sigma$ of the observed parameters.
We find that p-modes at radial orders of $n_p \approx 5-7$ are excited, having positive stability parameters (see Figure \ref{fig: etaplot}) that agree with the observed frequency range.

\section{Evolution and Discussion}\label{sec: evolution}
With an A-type subgiant primary star and a low-mass  ($\approx 0.2\,M_\odot$) evolved secondary star, KIC 12268220
might have a similar evolutionary history to the EL CVn system such as KIC 8262223 \citep{Guo2017} and KIC 7368103 \citep{Wang2019}. 
They are formed through the case-B evolution \citep{paczynski1971}, which leads to a mass exchange; the initial high-mass star evolves quickly to fill its Roche lobe and transfer its mass to the low-mass secondary. Recently, \citet{Chen2017} introduced the nonconservative stable mass transfer channel, which successfully explained the formation of the EL CVn and they also showed a grid of possible parameter space. 

To study the evolution of the KIC 12268220, we follow the method of \citet{Chen2017} and run a grid of models. However, solving the initial parameters from the current evolution stage is the inverse problem. It is not only time-consuming but also highly sensitive to the initial parameters. As discussed in \citet{Chen2017}, the mass transfer rate, angular momentum loss, and metallicity could have significant effects on the parameter space. Therefore, we run some theoretical models with different initial parameters and only display similar evolutionary tracks for some typical final WD masses.

\begin{figure}
    \centering
    \includegraphics[scale=0.6]{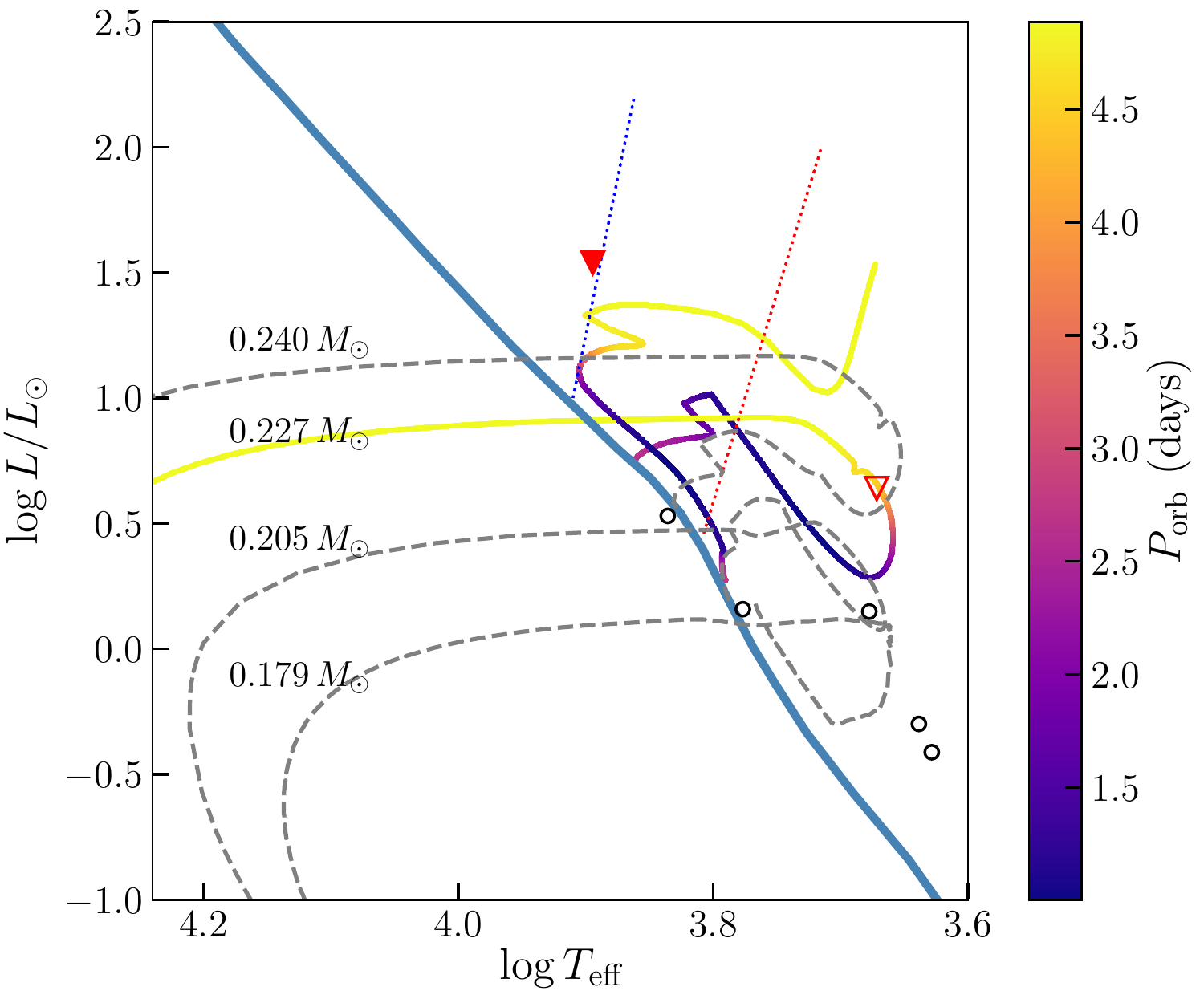}
    \caption{Evolutionary tracks in the Hertzsprung--Russell (HR) diagram. The primary and secondary star of KIC 12268220 are plotted as the filled and open red triangles respectively. Two lines in color indicate the typical model for primary and secondary components, and the orbital periods are color coded. The five cool R CMa-type objects are plotted in open black circles. Four evolutionary tracks with different WD masses are plotted, and the $0.179\,M_\odot$ track is adopted from \citet{Driebe1998}. The ZAMS is plotted as a thick steel-blue line. The $\delta$ Scuti instability strip is plotted by the dotted red and blue lines based on \citet{Xiong2016}.}
    \label{fig: evolution}
\end{figure}

We use the MESA evolution code with the Ritter \citep{Ritter1988} mass transfer scheme and a 50\% mass transfer rate; the initial metallicity is set to $ Z = 0.02 $ and the initial helium abundance is $ Y = 0.28 $. The gravitational wave radiation and magnetic braking are also switched on. In our models, we find some evolution tracks are similar to the KIC 12268220. For example, we choose the initial primary mass $M_{10} = 1.55\,M_\odot$, initial secondary mass $M_{20} = 1.23\,M_\odot$, and initial orbital period $P_\mathrm{orb0} = 2.75$ days. As shown in Figure \ref{fig: evolution}, the two lines in color indicate the evolutionary tracks for the primary and secondary stars respectively. This system ends up with the $M_1 = 1.89\,M_\odot$, $M_2 = 0.227\,M_\odot$, and orbital period $P_\mathrm{orb} = 4.87$ days. Although the model parameters and observed parameters of KIC 12268220 are slightly different, they are likely to experience a similar evolutionary process. Thus, we could use this model to study the evolutionary details.

From the evolutionary tracks shown in Figure \ref{fig: evolution}, because the mass transfer leads to the mass and radius decrease of the primary star, the most luminous point of the primary star after first evolving from the zero age main-sequence (ZAMS) indicates the onset of mass transfer; moreover, with the end of stable mass transfer, the orbital period keeps stable, so the ending point of the mass transfer is shown as the start of the yellow range.
After the typical L shape phase in the evolutionary track of the primary star, it evolves to a low-mass helium WD, with the secondary leaving its main sequence. 

Therefore, for KIC 12268220, as a semidetached system, according to the position of its secondary star on the evolutionary tracks of the model, we could infer it is about to or has just finished its mass transfer. Several studies also investigate some similar cool EL CVn candidates to KIC 12268220, including KIC 10661783 \citep{Southworth2011, Lehmann2013}, KIC 8262223 \citep{Guo2017}, KIC 7368103 \citep{Wang2019}, AS Eri \citep{Mkrtichian2004}, and R CMa \citep{Lehmann2018}. The parameters of their cool secondaries are listed in Table \ref{tab: cool-secondary}.

\begin{deluxetable}{ccccc}
\tablecaption{Table of known cool R CMa-type secondaries.\label{tab: cool-secondary}}
\tablehead{\colhead{Name}&\colhead{$M_2$}&\colhead{$R_2$}&\colhead{$T_\mathrm{eff2}$}&\colhead{$P_\mathrm{orb}$}\\
\colhead{}&\colhead{($M_\odot$)}&\colhead{($R_\odot$)}&\colhead{(K)}&\colhead{(days)}}
\startdata
KIC 10661783 & 0.20 & 1.12 & 5980 & 1.23\\
KIC 8262223 & 0.20 & 1.31 & 6849 & 1.61\\
KIC 7368103 & 0.21 & 1.75 & 4771 & 2.18\\
AS Eri & 0.21 & 1.15 & 4250 & 2.26\\
R CMa & 0.216 & 1.2 & 4350 & 1.13
\enddata
\end{deluxetable}

They are classified as the R CMa-type system, which is an Algol-type system with a low-mass ratio and short orbital period \citep{Budding2011}. Previous works believe they would most likely evolve into an EL CVn system \citep{Lee2018, Wang2019}. To compare with those cool progenitors of EL CVn, we also plot them on the HR diagram in Figure \ref{fig: evolution}. Some typical evolutionary tracks of different WD masses are also plotted as references. 
Since the $M_2$ of those R CMa-type systems are around 0.2--0.21$\, M_\odot$, we could estimate their evolutionary status from the relative position of those objects to the $0.205\, M_\odot$ track. The objects with relatively higher temperatures are more likely to start their L shape evolution; however, the lower temperature objects are still losing their masses during the mass transfer. After the mass transfer is finished, the final WD mass would be lower than the current $M_2$ and the orbital period would increase.

\begin{figure}
    \centering
    \includegraphics[scale=0.6]{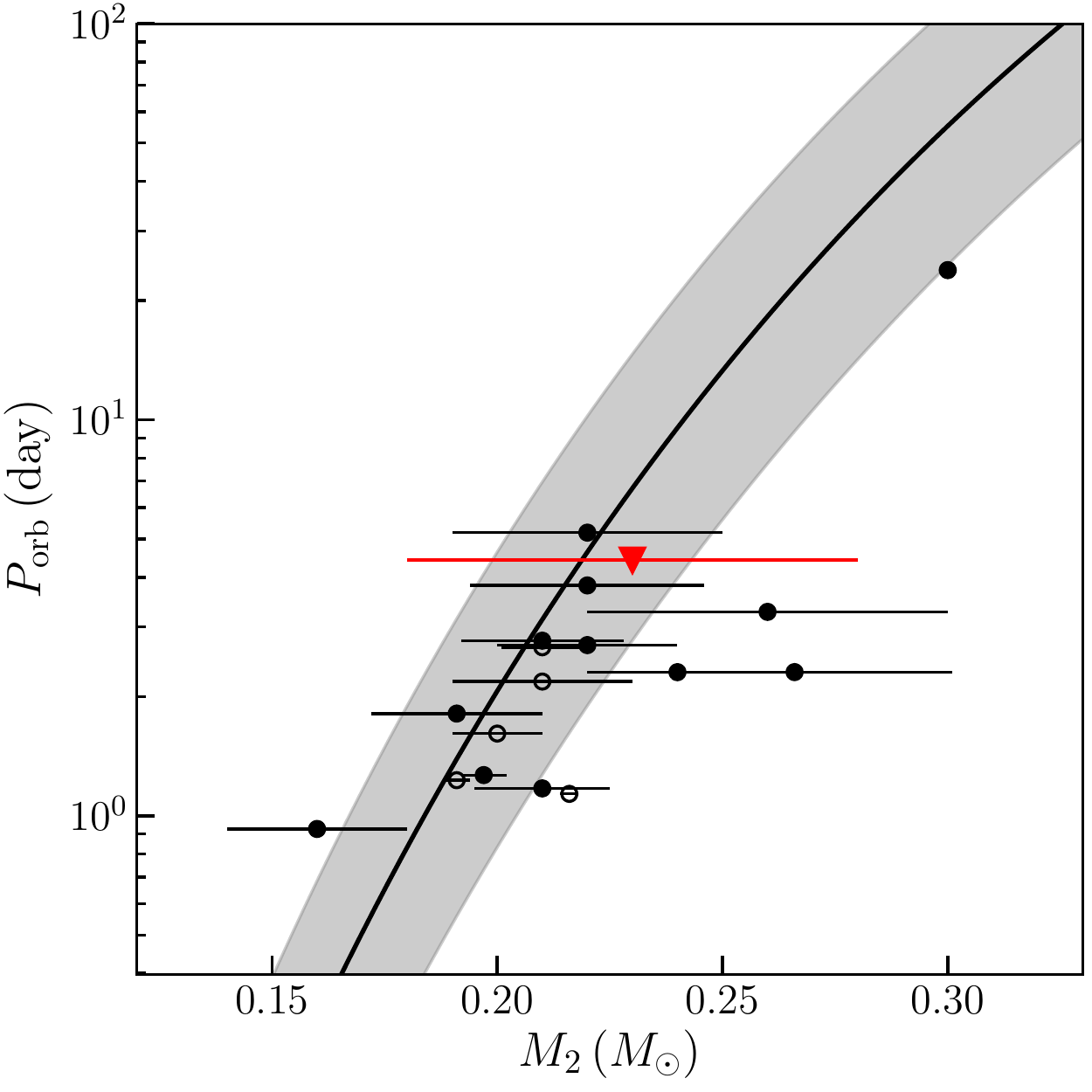}
    \caption{WD mass ($M_2$) versus the orbital period. The black line is adopted from \citet{Lin2011}, and the gray shaded region is the 10\% uncertainties of the $M_\mathrm{WD}$. The red triangle indicates the secondary component of KIC 12268220. The filled black circles are WDs discovered by Kepler and the open black circles are the five R CMa-type systems.}
    \label{fig: porb-mass}
\end{figure}

\citet{Chen2017} found a tight relation between the WD mass and the orbital period ($M_\mathrm{WD}$--$P$), which follows the formula of \citet{Lin2011}. Since this relationship is determined by the degenerated core mass-luminosity relation, it would be nearly stable no matter what the initial parameters of a binary are. 
We put KIC 12268220 on Figure \ref{fig: porb-mass} to check its $M_\mathrm{WD}$--$P$ relation. We also add the known WDs in the EL CVn systems found by Kepler \citep[collect from][]{Zhang2017} and the five cool R CMa objects to Figure \ref{fig: porb-mass}. It is clear that most of the WDs follow this relation and KIC 12268220 locates near the predicted line, although the error is relatively large.
From the $M_\mathrm{WD}$--$P$ results in Figure 11 of \citet{Chen2017}, most of the final products of their models locate on the left side of the predicted line except for a larger dispersion for $M_\mathrm{WD} \approx 0.23 $. However, unlike WDs, all the R CMa-type objects locate on the right side of the predicted line (red triangle), which could be confirmed by Figure 7 of \citet{Wang2019}. Here we explain this with the ongoing evolution for some of the R CMa-type systems, which means they would move to the upper left on the $M_\mathrm{WD}$--$P$ relation slightly after further evolution.

Thus, R CMa-type system might be a transition to the EL CVn system for low-mass and short orbital period ($M_\mathrm{WD} \lesssim 0.22\,M_\odot$, $P_\mathrm{orb} \lesssim 3$ days) objects;
similar mass but longer orbital period ($P_\mathrm{orb} \approx 4$ days) like KIC 12268220 would have slightly higher WD mass after evolved to be an EL CVn system.
Another possible effect on the evolution is the magnetic field \citep[e.g.,][]{Mestel1968, Rappaport1983}: a strong magnetic field would lead to more angular momentum loss. KIC 12268220 potentially possesses a strong magnetic field, which makes it an ideal target to study the role of the magnetic field during the evolution as an EL CVn precursor.

\section{Summary}\label{sec: summary}
In this paper, we investigate the eclipsing binary KIC 12268220 with the photometric and spectroscopic data. The spot features in the light curve suggests that KIC 12268220 has a strong magnetic activity. Combined with the atmospheric parameters, RV data, and Gaia-derived luminosity, the modeling of the light curve yields the fundamental parameters for both primary and secondary stars. The A-type primary star shows $\delta$ Scuti pulsations. We confirm the frequency range of unstable modes with the nonadiabatic theory. After running a grid of models following \citet{Chen2017}, similar to the R CMa, we suggest the low-mass secondary star is a precursor of helium WD.

\acknowledgments

We appreciate the referee for several suggestive
comments that significantly improved our article. We thank the help of Xinghao Chen, Gang Li, Namekata, Daniel Hey, Liang Wang, Yaguang Li, Changqing Luo and all the people who are interested in my poster at TESS Science Conference \uppercase\expandafter{\romannumeral1}.
We also thank the Xinglong Observational Astrophysics Training Workshop for the spectrum reduction and the Simulating Stars Summer School, 2018 in University of Chinese Academy of Sciences (UCAS) for the MESA technique.
We acknowledge support from the Chinese Academy of Sciences (grant XDB09000000), and from the National Science Foundation of China (NSFC, Nos. 11988101, 11425313, 11933004, 11903047).
New data presented here were observed with the 2.16\,m telescope at Xinglong Observation base in China.
We thank the support of the staff of the Xinglong 2.16\,m telescope. 
This work was partially supported by the Open Project Program of the Key Laboratory of Optical Astronomy, National Astronomical Observatories, Chinese Academy of Sciences.
This work was also partially supported by funding from the Center for Exoplanets and Habitable Worlds.  The Center for Exoplanets and Habitable Worlds is supported by the Pennsylvania State University, the Eberly College of Science, and the Pennsylvania Space Grant Consortium.
The paper includes data collected by the Kepler mission. Funding for the Kepler mission is provided by the NASA Science Mission Directorate. 
All of the Kepler data presented in this paper were obtained from the MAST. STScI is operated by the Association of Universities for Research in Astronomy, Inc., under NASA contract NAS5-26555. 
Support for MAST for non-HST data is provided by the NASA Office of Space Science via grant NNX09AF08G and by other grants and contracts. 

\vspace{5mm}


\software{numpy \citep{Walt2011}, 
          scipy \citep{Virtanen2020}, 
          matplotlib \citep{Hunter2007}, 
          pandas \citep{mckinney-proc-scipy-2010, reback2020pandas}, 
          astropy \citep{Collaboration2013,Collaboration2018a}, 
          IPython \citep{Perez2007},
          emcee \citep{ForemanMackey2013}
          }

\bibliography{12268220.bib}



\end{document}